\documentclass[12pt]{iopart}
\usepackage[dvips]{graphicx}

\begin{document}

\title[CTMRG and interacting SAW models]{A corner transfer matrix
renormalisation group investigation of the vertex-interacting 
self-avoiding walk model} 

\author{D P Foster and C Pinettes}
\address{Laboratoire de Physique Th\'eorique et 
Mod\'elisation (CNRS UMR 8089),
Universit\'e de Cergy-Pontoise,
5~Mail Gay-Lussac 95031 Cergy-Pontoise Cedex, France}

\begin{abstract}
A recently introduced  
extension of the Corner Transfer Matrix Renormalisation Group
(CTMRG) method useful for the study of self-avoiding walk type models
is presented in detail and applied to a class of interacting
self-avoiding walks due to Bl\"ote and Nienhuis.   This model displays
two different types of collapse transition depending on model
parameters. One is the standard $\theta$-point transition. The
other is found to give rise to a first-order collapse transition
despite being known to be in other respects critical.
\end{abstract}

\pacs{05.20.+q 36.20.-r 64.60.-i}
\submitto{\JPA}


\maketitle

\section{Introduction}

Self-avoiding walk models are of interest as models of polymers in
solution\cite{genbks} as well as realisations of $O(n)$ spin models in
the limit $n\to 0$\cite{degennes1}. 

The self-avoiding walk problem is a difficult problem numerically because
of the intrinsic long ranged nature of the problem; self-avoiding
walks are extended objects. The numerical methods most used to date
for the 
study of 
such models are Monte Carlo simulations\cite{landbind}, series
expansions\cite{gutt-domb} 
and transfer matrix calculations\cite{klein,enting,der,derh,ders,yeomv}. 

Monte Carlo simulations are hampered by  problems of trapping in the
low temperature, or equivalently large interaction, phases. These may
be relieved using the fluctuating bond method\cite{landbind,fluct-bnd}, 
which also enables the
investigation of the dense walk phases. This method has the drawback
of modifying the model in a way which is not always convenient,
particularly if frustration effects are of interest.
Current Monte Carlo
methods, such as the PERM method\cite{perm} and 
parallel tempering\cite{tvrow}, relieve these difficulties to some
extent, but at 
the cost of computational complexity.

Series expansions are limited by the lengths of walks that may be
enumerated. In practice these are limited to a few tens of lattice
steps. Very technical methods may be used to achieve longer series
expansions of the partition function (of the order of 60 lattice
steps), and in some cases the estimate
of the critical point is very impressive\cite{jg}. 
If correlation length
exponents are 
required, then the range of walk lengths open to enumeration are of
the order  
of 30 steps\cite{fot}. 
This becomes increasingly limiting as the complexity of
the model increases.

Transfer matrices have the advantage over Monte Carlo simulations of
giving numerically exact values of the free energy, correlation
lengths, polymer densities,etc., for strips of infinite length, but of
finite width. Finite size analysis methods then allow the extrapolation
of the 
finite size estimates.
The widths of lattice
accessible are limited: for the self-avoiding walk model widths up to
12 or 14 are possible (see for example \cite{blbatnien}), 
but this again drops as the complexity of the
model increases (see for example \cite{fs}). 
It often occurs that the number of data points one
may use in finite size analysis becomes dangerously small for 
finite size scaling techniques to be of use, particularly if
odd/even parity effects are present.

In this article we apply a recent extension of the corner transfer
matrix renormalisation group (CTMRG) method\cite{fp2003} to a class of
self-avoiding 
walks introduced by Bl\"ote and Nienhuis, which we shall refer to as
the vertex-interacting self-avoiding walk model, to reflect the fact
that the monomer-monomer interactions are added for doubly visited
sites (for a full description of the model see \Sref{moddef}).
We will show that the CTMRG method extends the possibilities of
transfer matrices to larger system sizes, at the cost of introducing a
phase space truncation scheme. The accuracy of the results is however
controllable through a parameter $m$ corresponding to the number of
states kept from one iteration to the next.

The other important result in this paper relates to the
vertex-interacting self-avoiding walk model. Whilst it has the
standard collapse transition as a special case (where the walk is
obliged to turn after each step), it is believed to have
a generic collapse transition in a different universality
class\cite{blbatnien}. We show that this collapse transition has a
first order character, despite the fact that there exists critical
exponent $\nu=12/23$ which has been exactly determined by Warnaar
\etal\cite{warnaar92}.

\section{The corner transfer matrix renormalisation method}

Transfer matrices, as a numerical method, have a major advantage
over Monte Carlo methods in that they sum over all possible
configurations, and so are not hampered by trapping effects
at low temperatures (or high densities in polymer problems) which
plague Monte Carlo simulations of models with competing interactions
(frustrated models).

The major inconvenience is the large amount of computer memory
required to store the matrices, limiting the maximum lattice width
 in practical calculations. This is alleviated in part
 by finite-size analysis methods which work well\cite{mpn76},
at least for the simpler
models, coupled with efficient extrapolation methods\cite{bst}.

In two-dimensional classical spin models, or
one-dimensional quantum
spin models, there exist iterative approximation schemes
which, by successive pruning of phase space,
permit calculations for
much larger lattice widths for given
computer resources.
These methods are the Density Matrix
Renormalisation Group method (DMRG)\cite{white,peschel99} 
and the Corner Transfer Matrix
Renormalisation Group method (CTMRG)\cite{nishino}.
These methods have little
recourse to the renormalisation group, and are better thought
of as iterative matrix approximation schemes. The basic idea is
that the (approximate) transfer matrix for a lattice of size $2N$ is
calculated from the transfer matrix for a lattice of size $2N-2$
by inserting extra lattice sites. A change of basis is then
performed, such that the transfer matrix may be projected onto a
smaller basis, with the smallest possible loss of information.

In a recent paper we showed how the CTMRG method, first introduced for
classical spin systems by Nishino\cite{nishino}, 
may be applied to self-avoiding walk
type models\cite{fp2003}. 
This requires a non-trivial modification of the method
since the  iteration scheme requires all interactions to be local.
The self-avoiding walk problem, by its very nature, is non-local. In the
remainder of this section the details of the method, absent in
reference\cite{fp2003}, are provided. In particular
we show how this non-locality may be
overcome, and describe the implementation of CTMRG for this class of
problem. 

\subsection[SAW and Vertex models]{The self-avoiding walk, the $O(n)$
model and vertex
models with complex weights}\label{moddef}

It is known that  spin models with $O(n)$ invariant spins have
high temperature expansions consisting of closed loops.
The self-avoiding loop (SAL) model is related to the high
temperature expansion of the $O(n=0)$ model\cite{degennes1}. 
In a similar way, the
high temperature expansion of the susceptibility leads to a model
with one open path, and a gas of loops\cite{jmy}. The $n\to 0$ limit
eliminates the loops to leave a self-avoiding walk. The partition
function for a self-avoiding walk is thus equivalent to the
susceptibility of an $O(n=0)$ spin model.

The connection between the $O(n)$ class of models and better known
models is as follows: $n=1$ corresponds to the Ising model, $n=2$
to the XY model and $n=3$ to the classical Heisenberg model. The
partition functions of these models are given by:
\begin{equation}\label{onpart}
{\cal Z}=\sum_{\{\vec{s}_i\}} \exp\left(\frac{1}{2}\beta
J\sum_{\langle i,j \rangle} \vec{s}_i\cdot\vec{s}_j\right),
\end{equation}
where $\langle i,j\rangle$ refers to a sum over nearest
neighbour spins.
 Whilst the high-temperature expansion of
the Ising model is relatively easy, the expansion for other values
of $n$ is a little cumbersome.
In what follows we shall consider a slightly different $O(n)$
invariant model, due to Nienhuis\cite{nienhuis82,nienhuis87}, 
with the partition
function given by
\begin{equation}\label{nonpart}
{\cal Z}_{O(n)}=\sum_{\{s^{\alpha}_i\}} \prod_{\langle i,j\rangle}
\left(1+K\sum_{\alpha=1}^n s^\alpha_i s^\alpha_j\right),
\end{equation}
where the spins, $s^\alpha_i=\pm 1$, are placed on the lattice bonds.
A diagrammatic
expansion of Equation~\ref{nonpart} follows if we identify the $1$
as the weight of an empty bond between the sites $i$ and $j$ and
the $K$ as the weight of an occupied bond. Since $\langle
s^\alpha_i\rangle=0$ and $\left(s^\alpha_i\right)^2=1$, the only
terms in the expansion of Equation~\ref{nonpart} which survive the
sum over $\{s_i^\alpha\}$ are those terms consisting of closed
loops where the `colour' $\alpha$ is conserved around each
loop. Different loops may have different `colours'. The sum over
$\alpha$ gives a factor $n$ for each loop. An equivalent
expression for Equation~\ref{nonpart} may now be given in terms of
loop graphs $\cal G$:
\begin{equation}\label{gpart}
{\cal Z}_{O(n)}=\sum_{{\cal G}} n^{l({\cal G})} K^{b({\cal G})},
\end{equation}
where $l$ is the number of  loops and $b$ is the number
of occupied bonds. The partition for self-avoiding loops is then
just
\begin{equation}
{\cal Z}_{SAL}=\lim_{n\to 0} \frac{1}{n} {\cal Z}_{O(n)}.
\end{equation}

\begin{figure}
\caption{The 19 allowed vertices in the Nienhuis $O(n)$ class of
models, with the weights for the vertex-interacting 
self-avoiding 
walk model
($\theta=\pi/2$). $K$ is the step fugacity and
$\tau=\exp(-\beta\varepsilon)$ is the Boltzmann weight associated with
non-intersecting site interactions.}\label{verts}
\begin{center}
\includegraphics[width=11cm,angle=-90,clip]{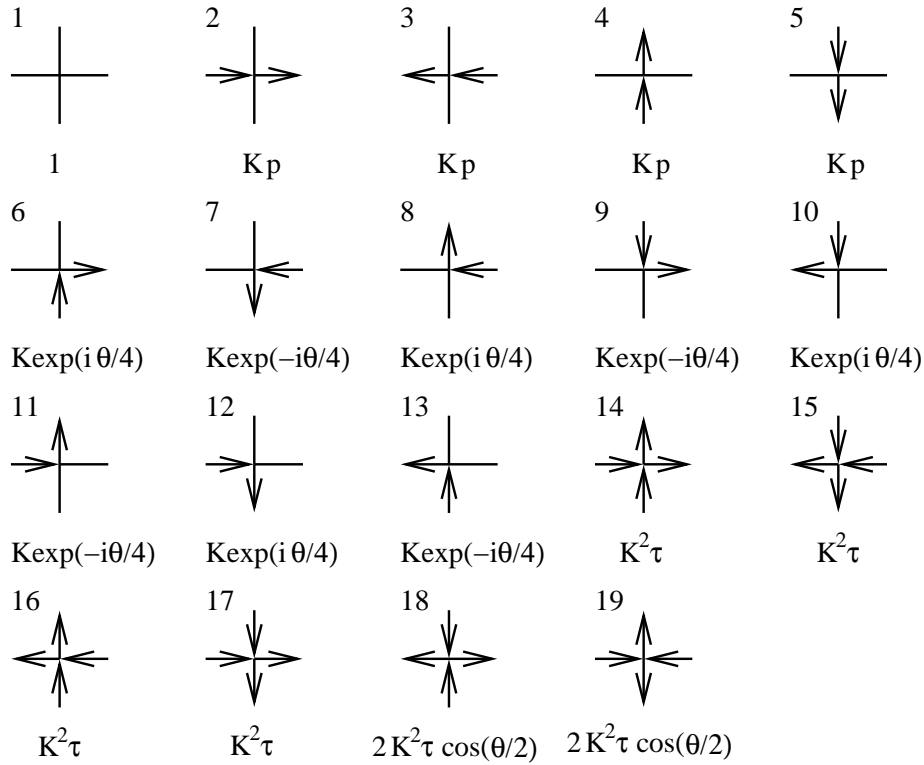}
\end{center}
\end{figure}

The parameter  $n$ is now a fugacity controlling the number of
loops, and in this representation need no longer be taken as an
integer. This fugacity corresponds to a long-range interaction,
since the loops may be of any size. This non-locality is
undesirable for our purposes. We would like to express $n$ as a
product over local weights despite the variable size of the loops.
The only real numbers with this property are $0$ and 1. This leads
us to introduce complex local weights,
$w_i$.  If $|w_i|=1$, the
product of the $w_i$ around a loop will also be of modulus 1. The
final weight must, however, be real. This may be achieved if for
every complex weight, we also have
its complex conjugate. Each loop may
be followed clockwise or anticlockwise.
The loops in
Equation~\ref{gpart} are
 not oriented, but may be oriented
by associating $2^{l({\cal G})}$ oriented graphs with each
non-oriented graph. The loop fugacity is not required to be the same
for the two orientations of the loop, and so a fugacity $n_+$ ($n_-$)
is associated with (anti)clockwise oriented loops.
The partition function
may then be written\cite{nienhuis82,nienhuis87}
\begin{eqnarray}
{\cal Z}_{O(n)}&=&\sum_{\cal G} (n_+ + n_-)^l K^b,\\
 &=&\sum_{{\cal G}^{\prime}}n_+^{l_+} n_-^{l_-} K^b
\end{eqnarray}
where ${\cal G}^\prime$ is the set of oriented loop graphs and $l_+$ ($l_-$)
is the number of (anti)clockwise oriented loops.
On the square lattice there must
be four more corners with one orientation, compared to the other
orientation, in order to close a loop. Using this fact we may 
 set $w_i=\exp(i
\theta/4)$ for a clockwise corner, $w_i=\exp(-i\theta/4)$ for an
anticlockwise corner and $w_i=1$ otherwise.
This leads to $n_+=\exp(i\theta)$ and $n_-=\exp(-i\theta)$, 
and hence $n=2\cos(\theta)$. The
oriented loop factor has now been broken up into local weights and
the partition function may be rewritten in terms of a vertex
model\cite{baxter86,nienhuis90}:
\begin{equation}\label{vertpart}
{\cal Z}=\sum_{{\cal G}^\prime} \prod_i v_i
\end{equation}
where $v_i$ is the weight of the vertex found at site $i$. The
possible vertices are shown in Figure~\ref{verts}. The derivation
given here is only for the simplest case, but we may freely change
the weights of the vertex configurations in order to generate
different interactions in the original model. Notably, to have
exactly the standard self-avoiding walk model it is necessary to
set the weights of vertices 14 to 19 to zero, and set $n=0$, or
equivalently set $\theta=\pi/2$.

A vertex model may be thought of as a spin model where the spins
live on the bonds of the lattice. Each spin has three states
corresponding to the empty lattice bond, a bond oriented in the
positive $x$ ($y$) direction and a bond oriented in the negative
$x$ ($y$) direction. The vertex weights may be written as a vertex
function $W(\sigma_i,\sigma_j,\sigma_k,\sigma_l)$, where the
tensor elements correspond to the vertex weights of
Figure~\ref{verts}.
The vertex function was defined using the following convention: 
\begin{equation}
\sigma=\cases{
1 & {\rm if the
bond has an arrow to the right or upwards,}\\ 
0 & \textrm{if the bond has no arrow,}\\
-1 & \textrm{if the bond has an arrow to the left or down.}
}
\end{equation}

\subsection{The vertex-interacting self-avoiding walk model}

In this section we apply the CTMRG method explicitly to the vertex-interacting
self-avoiding walk model introduced by Bl\"ote and
Nienhuis\cite{nienhuis90,nienon}.   
This model consists of a random
walk which is allowed 
to collide at a site, but is not allowed to cross or occupy a bond
more than once. An interaction energy $\varepsilon<0$ is introduced
for each collision, which replaces the attractive nearest-neighbour
interaction present in the standard $\Theta$-point model. A step fugacity,
$K$, is introduced in order to control the average length of the
walk. A stiffness is also introduced by weighting sites sitting on a
bend in the walk differently from sites sitting on straight portions
of the walk. Following the convention of Bl\"ote and Nienhuis\cite{nienon}, we
choose to add a weight $p$ to the straight segments of the walk. The
choice $p=0$ then corresponds to a walk in which straight segments are
eliminated (the walk is obliged to turn through a right angle after
each step). The partition function for the model is then given by
\begin{equation}
{\cal Z}=\sum_{\textrm{walks}} (Kp)^{N_s} p^{-N_c} \tau^{N_I},
\end{equation}
where $\tau=\exp(-\beta\varepsilon)$, $N_s$ is the length of the walk
(number of steps), 
$N_I$ is the number of doubly
occupied sites and $N_c$  is the number of corners. 

\subsection[CTMRG]{The corner transfer matrix renormalisation group
method}\label{CTMRGsec}

\begin{figure}
\caption{A schematic representation of the two-dimensional lattice
split into four quarters, represented by the four corner matrices
{\bf A, B, C} and {\bf D}.}\label{corner}
\begin{center}
\includegraphics[width=11cm,angle=-90,clip]{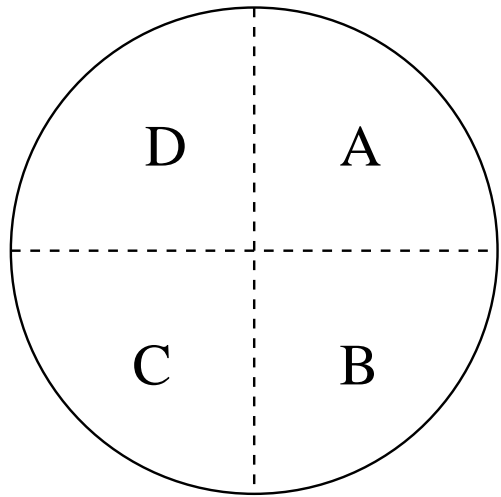}
\end{center}
\end{figure}

Following Baxter\cite{baxter68,baxter78,baxter}, 
the partition function of a two-dimensional
lattice model may be written in terms of the product of four
matrices representing the four quarters of the lattice, see
Figure~\ref{corner}.  The inputs and outputs of the matrices are
the configurations at the seams of the four quarters. These
matrices are known as corner transfer matrices. In general the
four matrices are different, but may often be related by lattice
symmetries. For our model the four matrices are the same up to a
complex conjugation operation.

\begin{figure}
\caption{Figure (A) shows the initial vertex lattice. The 3-state
spins defined in the vertex model are
represented by circles. The black spins
are summed over, taking into account the boundary conditions (fixed
or free), to give the prototype system used in the CTMRG method,
shown in (B). The 
squares represent the $m$-state spins used in the CTMRG method. $C_N$
is an $m\times m$ corner transfer matrix representing one quarter of
the lattice 
(see text). The asterisk represents complex conjugation.
}\label{proto}
\begin{center}
\includegraphics[width=12cm,angle=-90,clip]{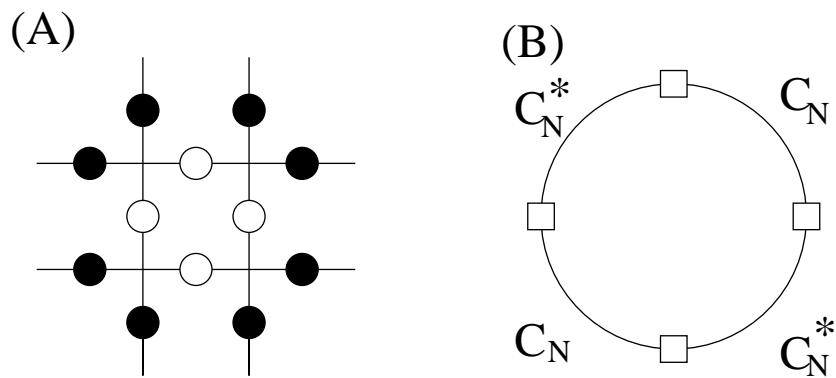}
\end{center}
\end{figure}

 In general it is difficult to explicitly calculate these
matrices for systems with a large number of sites. This is where
the CTMRG method comes in\cite{fp2003,nishino};
the matrices for larger lattices are
calculated iteratively from smaller lattices.  This is done as
follows. A prototype system consisting of a small number of spins,
four in the case of a vertex model, is set up exactly,
see Figure~\ref{proto}. At
each iteration the system is enlarged by adding spins, and this
enlarged system is then projected back onto the prototype system
in some optimal way, so as to minimise the loss of information.
The spins of the prototype system have $m$ states, where $m$
determines the amount of information which may be carried forwards
at each iteration.

\begin{figure}
\caption{The matrix $C^\prime_{N+1}$ is constructed by adding the
vertex functions $P_N$ and $W$ to $C_N$ as shown. The asterisk
represents complex
conjugation. Circles
represent the 3-state spin variables defined by the original
model, and the squares represent the $m$-state spins defining the
CTMRG prototype system. The black spins are summed over, leaving
the white spins as the indices of
$C^\prime_{N+1}$. $C^\prime_{N+1}$ is a $3m\times 3m$ matrix which
must then be projected in an optimal way to give the new ($m\times m$)
corner transfer matrix $C_{N+1}$. 
}\label{cornerb}
\begin{center}
\includegraphics[width=11cm,angle=-90,clip]{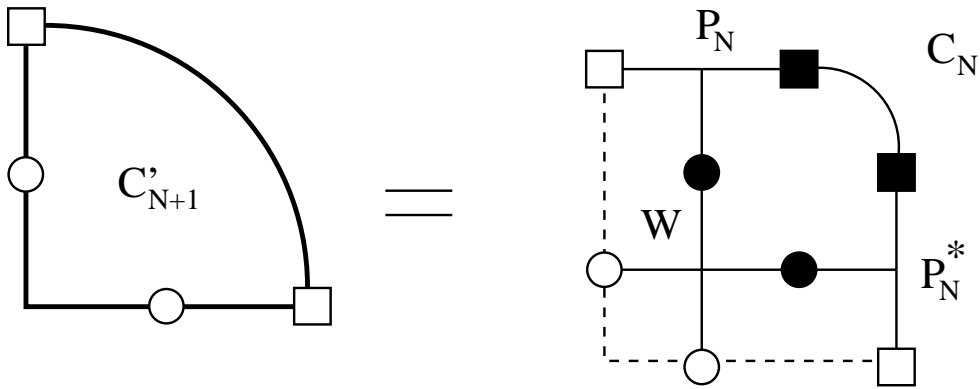}
\end{center}
\end{figure}

In addition to the vertex function
$W(\sigma_i,\sigma_j,\sigma_k,\sigma_l)$, we also introduce the
vertex functions $P_N(\sigma_i,\xi_j,\xi_k)$ and
$C_N(\xi_i,\xi_j)$ where the $\xi_i$ are the $m$-state spins of
the prototype system. The matrix $C_N$ corresponds to the estimate
of the corner transfer matrix after $N$ iterations. The partition
function for a $2N\times 2N$ lattice is given by
\begin{equation}
Z_N={\rm Tr} \left(C_N C_N^*\right)^2,
\end{equation}
where the $*$ denotes  complex conjugation.
The lattice is enlarged by introducing additional vertices, as
shown in Figure~\ref{cornerb}. A corner matrix for the enlarged
lattice is given by
\begin{equation}
C_{N+1}^{\prime}=W\bullet
P_N\bullet P_N^* \bullet C_N,
\end{equation}
where $\bullet$ indicates that the appropriate contractions over
indices are performed. In a similar way we introduce an enlarged
vertex function
\begin{equation}
P_{N+1}^{\prime}=W\bullet P_N.
\end{equation}

By considering carefully the symmetry of the vertex function $W$ it
may be shown that the partition function for the enlarged lattice
$2(N+1)\times 2(N+1)$  
is
given by
\begin{equation}
Z_{N+1}^\prime={\rm Tr} (C_{N+1}^{\prime} C_{N+1}^{\prime *})^2.
\end{equation}

We now wish to project the enlarged model, defined through
$C_{N+1}^\prime$ 
and $P_{N+1}^\prime$,  
back onto the smaller
prototype system, giving $P_{N+1}$ and $C_{N+1}$.

Since the configuration space is smaller after projection, some
information is lost. The trick is to minimise this information
loss. This is done by performing a change of basis in some optimal
manner. The partition function $Z_{N+1}^\prime$ is invariant by
any change of basis, but the partial trace, giving $Z_{N+1}$ after
the change of basis, is not. For a model with real weights it is
clear that the optimal choice will minimise
$\varepsilon=|Z^\prime_{N+1}-Z_{N+1}|$. This is done by
diagonalising the  matrix $\tilde{z}=(C^\prime_{N+1} C^{\prime
*}_{N+1})^2$ and 
keeping the basis vectors corresponding to the largest
eigenvalues. In a model with real weights $\tilde{z}$ is
just the density
matrix, up to a constant. 
In the case considered here the eigenvalues of $\tilde{z}$ are
complex, coming in conjugate pairs. Minimising $\varepsilon$
corresponds 
to taking the basis vectors
for the eigenvalues with the largest real parts, since
the imaginary parts cancel. This may well involve  dropping 
terms contributing to $Z^{\prime}_{N+1}$ with larger moduli than
other terms which are kept. It is not clear how this effect
propagates with the number of iterations. Another choice is to
keep the terms contributing to $Z^\prime_{N+1}$ with the largest
absolute values. This is found to give better results in practice.
To complete the iteration scheme, we must also calculate
$P_{N+1}$. This is done by projecting $P^\prime_{N+1}$ onto
$P_{N+1}$ using the same basis vectors. As this iterative process
is repeated, the partition function and other
thermodynamic quantities are calculated for larger and larger
lattices. The value of $m$ defines the size of the configurational
space kept from one iteration to the next. The larger the value of
$m$, the better the approximation. 

The iteration scheme not only gives an approximation of the partition
function for a square system, but also gives an approximation for the
transfer matrix from which an infinite strip may be constructed. The
transfer matrix is given directly in terms of the vertex functions
$P_N$ and $W$ as
\begin{equation}\label{tm1}
{\cal T}=P_N\bullet W\bullet W \bullet P^*_N
\end{equation}
for even lattice widths (width $L=2N+2$) and
\begin{equation}\label{tm2}
{\cal T}=P_N\bullet W \bullet P^*_N
\end{equation}
for odd lattice widths (width $L=2N+1$). 
Free energies and correlation lengths for
strips may then also be calculated, and the arsenal of finite-size
scaling methods available for strips may be used. CTMRG is most
efficient, however, when it is used to calculate directly  one-point
functions, such as the density, and this is how we use
it in this article.

The density is most simply calculated by selecting one of the central
lattice bonds. The density $\rho$ is then
given by $\rho=\langle|\sigma|\rangle$, where $\sigma$ is the
value of the spin on the selected bond. In terms of the corner
transfer matrices,
\begin{equation}
\rho_L=\frac{{\rm Tr}( \Lambda  C_N^{\prime} C_N^{\prime
*} 
C_N^{\prime}  C_N^{\prime*})}{{\rm Tr}( C_N^{\prime}
C_N^{\prime *} C_N^{\prime}C_N^{\prime *})}
\end{equation}
where $\Lambda$ is a matrix whose elements take the value $|\sigma|$
and $L=2N$ is the linear dimension of the lattice. 

\begin{figure}
\caption{Variation of the density $\rho$ for the self-avoiding walk
model ($p=1, \tau=0$) calculated using CTMRG as a function of the
linear size $L=2N$ of 
the lattice  for different values of $m$.}\label{sawm}
\begin{center}
\includegraphics[width=10cm,clip]{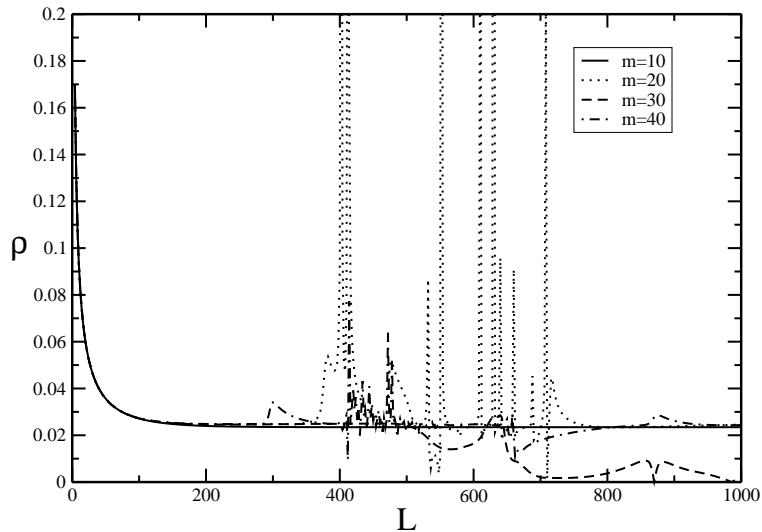}
\end{center}
\end{figure}

In spin models
the one-point functions are often calculated by the infinite lattice
method\cite{nishino}, where the CTMRG method is iterated for a fixed
value of $m$ 
until the quantity of interest (magnetisation or energy per spin) no
longer changes. This occurs because the number of eigenstates kept at
each iteration is no longer sufficient to propagate the two point 
correlations throughout the system. The limiting value of the one
point correlation function may then be studied as a function of
$m$. In such a manner the magnetisation or energy per site may be
estimated for the infinite system directly. This method relies 
on the rather
nice way in which CTMRG breaks down for spin
models. Unfortunately the complex nature of the vertex models at the heart
of our calculations gives rise to more complicated behaviour as $m$ is
increased. \Fref{sawm} shows the density of
the self-avoiding walk model ($p=1$, $\tau=0$ and $n=0$, or
equivalently $\theta=\pi/2$) as a
function of $L$ for different
values of $m$. It is clearly seen that there is no limiting value for
the density as $L$ is increased. Once the value 
of $m$ is insufficient, $\rho$ has a chaotic oscillatory behaviour with
$L$.
Instead of fixing $m$ and varying $L$, we choose to do the
opposite. Taking the density as an example, we define $\rho_L(m)$ as
the density for a system of linear dimension $L$ for a given value of
$m$. The limiting value of $\rho_L(m)$
is calculated by  increasing the value of $m$ for fixed $L$ until 
the required accuracy for $\rho_L$ is reached. This is then studied as
a function of $L$ using finite-size scaling methods, described below.

\section{Results}

\subsection{The vertex-interacting model with $p=0$}\label{p=0sec}

\begin{figure}
\caption{The schematic phase diagram for the 
vertex-interacting model with $p=0$. For $p=0$ the vertex-interacting 
model has the same schematic phase diagram as the $\Theta$-point
model, and the 
same critical behaviour. The solid line corresponds to a critical
phase transition in the self-avoiding walk universality class. The
dashed line corresponds to a first-order transition. The two types of
phase transition are separated by a tricritical point
$(K_{tc}=1/2,\tau_{tc}=2)$.}\label{thetapd} 
\begin{center}
\includegraphics[width=10cm,angle=-90]{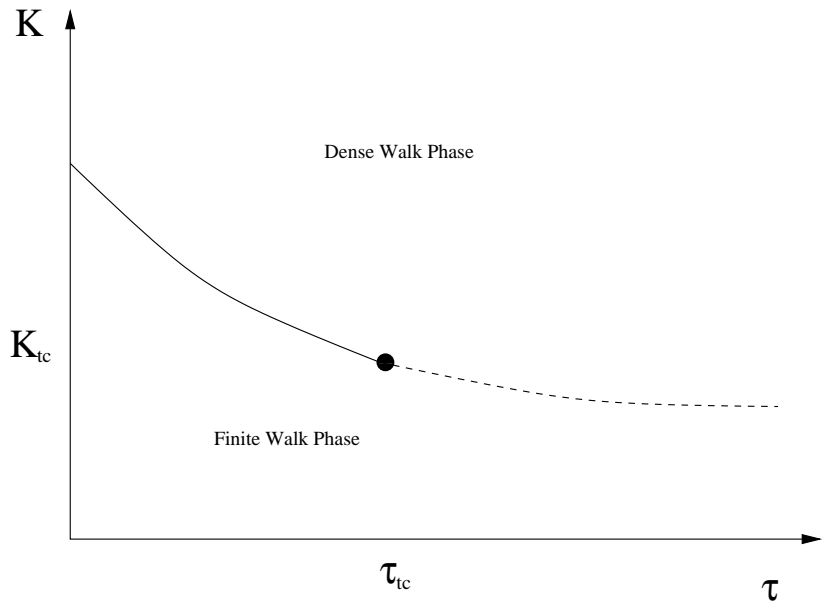}
\end{center}
\end{figure}

When $p=0$ the model is expected to have a critical behaviour
equivalent to the standard $\Theta$-point model\cite{blbatnien}; for
small enough 
$\tau$ the average length of the walk diverges as a critical value of
the step fugacity, $K_c$, is reached. At $K_c$ the density is still
zero and the gyration radius, $R_G$, (or any other characteristic measure of
the size of the walk) diverges as $K\to K_c^{-}$ following a power law
defining the correlation length exponent $\nu$:
\begin{equation}
\xi\sim (K_c-K)^{-\nu}.
\end{equation}
For $\tau$ smaller than some tricritical value, $\tau_{tc}$, the walk is in
the same universality class as the usual self-avoiding walk, and
$\nu=3/4$. This behaviour terminates at the tricritical point
$(K_{tc},\tau_{tc})$, 
where $\nu=4/7$ as for the standard $\Theta$-point. For values of
$\tau>\tau_{tc}$ the length of the walk diverges discontinuously at a
first 
order transition. The phase diagram of the $p=0$ vertex-interacting model
is qualitatively the same as for the $\Theta$ point\cite{blbatnien},
and is 
shown schematically in  \Fref{thetapd}. Unlike for the standard
$\Theta$-point model, the location of the tricritical point is known
exactly to be $(K_{tc}=1/2,\tau_{tc}=2)$\cite{blbatnien}.

The idea is now to study this model using the CTMRG method
outlined. As mentioned in \sref{CTMRGsec}, CTMRG gives as 
a byproduct an approximation of the
transfer matrix for a strip of width $L$ through 
\eref{tm1} or \eref{tm2}, depending on the parity of L, from which the
phase diagram and critical exponent $\nu$ could be calculated using
the phenomenological renormalisation group due to
Nightingale\cite{mpn76}. This turns out, at least for the case at hand, not to be
the most efficient way of approaching the problem, due to the
additional computational effort required to calculate the relevant
eigenvalues of $\cal T$. The CTMRG method is far more efficient for
the calculation of one-point functions, such as the bond density,
which may be calculated in the middle of the system. With this in mind
we use a phenomenological renormalisation scheme based directly on the
scaling behaviour of density.

The scaling ansatz states that the singular part of the free energy
is a homogeneous function of its arguments\cite{stanley}. This leads
to the following finite size scaling expression:
\begin{equation}
f_s(K,L)=L^{-d}\tilde{f}(|K-K_c|L^{1/\nu}),
\end{equation}
where $d$ is the spatial dimension, here $d=2$.
Taking the first derivative,
\begin{equation}\label{scalingfn}
\rho_s(K,L)=L^{1/\nu-2}\tilde{\rho}(|K-K_c|L^{1/\nu}),
\end{equation}

In general 
\begin{equation}\label{gendens}
\rho(K,L)=\rho_{\infty}(K)+\rho_s(K,L),
\end{equation}
and so it is
necessary to know the value of $\rho_{\infty}(K)$ in order to
exploit fully the scaling behaviour. The density
of a finite walk on an infinite lattice is zero, and so
$\rho_{\infty}(K<K_c)=0$. 
This enables the setting up of a phenomenological
renormalisation group method using  the function
\begin{equation}\label{oldsfn}
\varphi_{L,L^\prime}(K) =
\frac{\log\left(\rho_s(K,L)/\rho_s(K,L^\prime)\right)}{\log\left(L/L^\prime
\right)}.
\end{equation}
Using \eref{scalingfn} we find that $\varphi_{L,L^\prime}=1/\nu-2$
independently of $L$ and $L^\prime$. Naturally there are additional
finite size corrections which should be taken into account, but the
conclusion is that if the function $\varphi_{L,L^\prime}(K)$ is
plotted for various values of $L$ and $L^\prime$ then it will converge
to a fixed point given by $\varphi(K_c)=1/\nu-2$. In what follows we
have set $L^\prime=L-2$ and looked for solutions of the equation
\begin{equation}\label{ssol}
\varphi_{L,L-2}(K^L_c)=\varphi_{L-2,L-4}(K^{L}_c).
\end{equation}
If such solutions exist then $K_c=\lim_{L\to\infty}K_c^L$ and
$\nu=\lim_{L\to\infty}1/(2+\varphi_{L,L-2}(K^L_c))$.



\begin{figure}
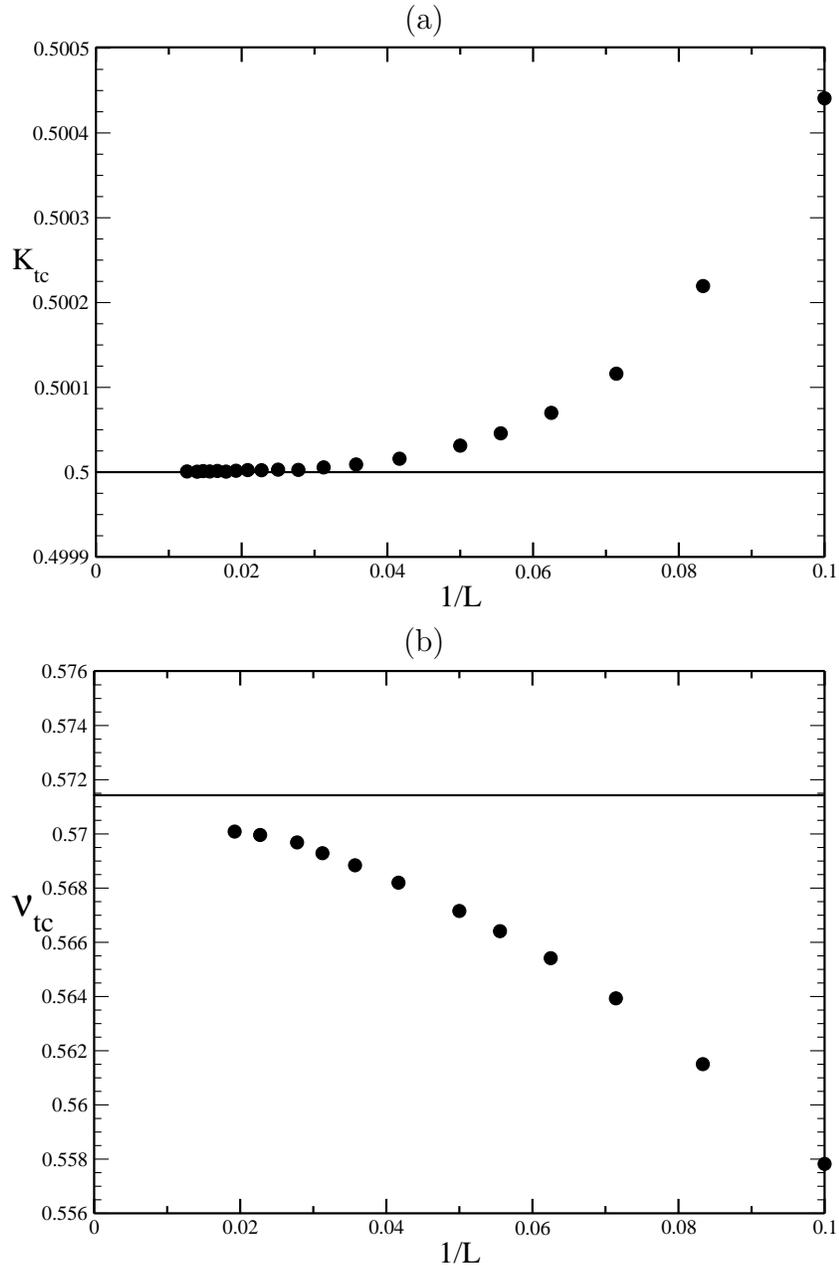

\caption{Estimates of the tricritical values  $K_{tc}$ and
$\nu_{tc}$ calculated for $p=0$ and
$\tau=\tau_{tc}=2$ using \eref{ssol}, plotted as a function of the
linear size $L$. There is a clear
convergence to the expected exact values $K_{tc}=1/2$ and
$\nu_{tc}=\nu_{\theta}=4/7$.}\label{nthp}
\begin{center}
(a)

\includegraphics[width=11cm,clip]{nthtc}

(b)

\includegraphics[width=11cm,clip]{nthnu}
\end{center}
\end{figure}

\begin{figure}
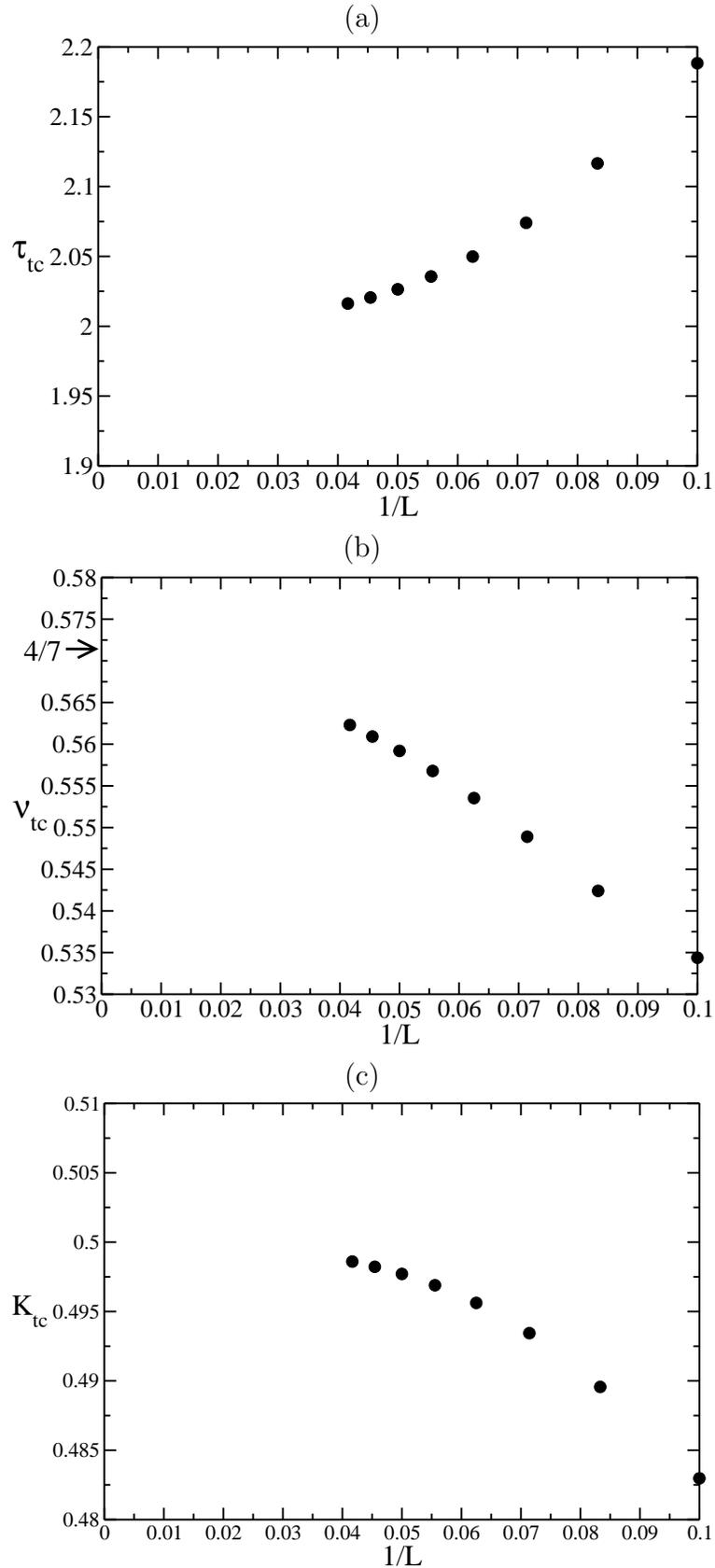

\caption{Estimates of the tricritical values (a) $\tau_{tc}$, (b)
$\nu_{tc}$ and (c) $K_{tc}$ 
calculated from the intersections of $\nu_L(\tau)$ for
$p=0$. There is a clear convergence to the exact values are
$\tau_{tc}=2$, $\nu_{tc}=4/7$ and  
$K_{tc}=1/2$}\label{theta_est}  
\begin{center}
(a)

\includegraphics[width=10cm,clip]{thetatc}

(b)

\includegraphics[width=10cm,clip]{thetanu}

(c)

\includegraphics[width=10cm,clip]{thetakc}
\end{center}
\end{figure}

Preliminary results using CTMRG were recently reported in reference
\cite{fp2003}; notably the density profiles and 
values for $K_c$ and $\nu$ are given for various values of $\tau$ with
$p=0$.
When $p=0$, 
the vertex-interacting $\Theta$ point is known to occur exactly at
$K_{tc}=1/2$ and $\tau_{tc}=2$.  In \Fref{nthp} we present results 
using this scheme  where we have fixed  $p=0$ and 
$\tau=\tau_{tc}=2$. The results 
converge rather quickly to the exact values, $K_c=1/2$ and
$\nu_{\Theta}=4/7$. 
Whilst here the tricritical values of $K$, $\tau$
and $\nu$ are known exactly, this is in general not the case. It
is interesting to know how well the method works 
with no external input. As we pass through $\tau_{tc}$ the value of $\nu$
changes from $\nu_{\rm SAW}=3/4$ for $\tau<\tau_{tc}$ to
$\nu_{\Theta}=4/7$ for 
$\tau=\tau_{tc}$ and on to $\nu=1/2$ for $\tau>\tau_{tc}$. This last value
is simply a reflection of the fact that the density is nonzero, and so
\mbox{$R_G\sim N^{1/2}$}. Clearly the finite size estimates, $\nu_L(\tau)$,
will be continuous functions of $\tau$, tending to the step function
only in the infinite lattice limit. If the different $\nu_L(\tau)$
cross, then these 
intersections necessarily tend, in the limit of an infinite system, to
the correct values:  $\nu_{tc}=\nu_\theta$ and
$\tau_{tc}$. This is a standard method used to determine
numerically the location, for example, of the $\Theta$ point in the
standard model (see for example \cite{der,derh,ders}). 
The estimates derived from this
method are shown in \Fref{theta_est} plotted as a function of
$1/L$. The estimates converge convincingly to the exact values
$\tau_{tc}=2$, $K_{tc}=1/2$ and $\nu_{tc}=4/7$.

\subsection{The vertex-interacting model with $p=1$}

What is interesting about the vertex-interacting model is that the phase
diagram is very different when $p\ne 0$\cite{nien}.
 Na\"{\i}vely one would expect
the $p=1$ model to also be equivalent to the $\Theta$-point model;
the model consists of a walk which is essentially
self-avoiding with short-ranged attractive interactions. This is
in fact not the case; an additional transition arises in the dense walk
phase and the nature of the transition at the (now) multicritical
point changes. There are no exact results for the
location of the multicritical point $(K_{mc},\tau_{mc})$ as was the
case for $p=0$. There exist, however,  exact predictions for the
universality classes of the different phase transitions. The
low-density phase transition is expected to be in the self-avoiding
walk universality class (with $\nu=3/4$)\cite{nienon,nien}.
The high-density critical line is expected to be in 
the Ising universality
class ($\nu=1$)\cite{nienon,nien}. The value of $\nu_{mc}$ at the multicritical point
has been found from an exact result for the 19 vertex model with complex
weights used in our calculations. It is predicted that
$\nu_{mc}=12/23$\cite{warnaar92}.  
In the remainder of this section we calculate the phase diagram
and compare  critical exponent estimates calculated from the CTMRG
method with the known results. 

\begin{figure}
\caption{The density of the walk on the lattice for $p=1$ and $\tau=1$
as a function of $K$, clearly showing the existence of two phase
transitions.}\label{p=1dens} 
\begin{center}
\includegraphics[width=12cm,clip]{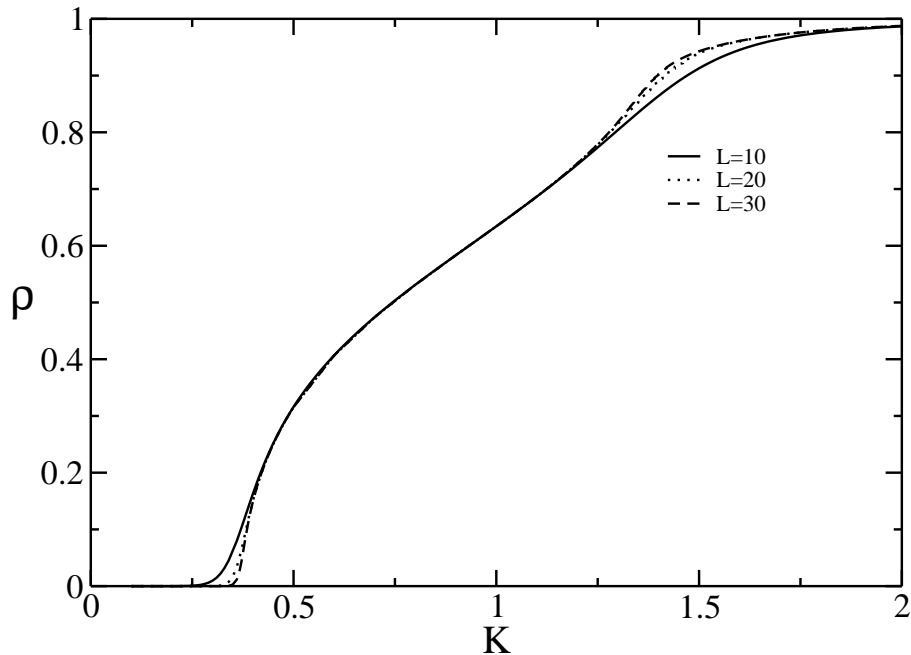}
\end{center}
\end{figure}

In \fref{p=1dens} the density is plotted as a function of $K$ for
$p=1$ and $\tau=1$. The presence of two phase transitions may clearly
be seen, the first between the zero-density phase and a dense walk
phase, and the second between two dense walk phases. 

For the first transition, as $K$ is increased, it is
possible to use the scaling arguments given in \sref{p=0sec} since
the density in the thermodynamic limit is zero. For the second
transition, however, the density is non-zero, and a priori unknown. To
apply the scaling arguments it is necessary to estimate the
density in the infinite lattice limit,
$\rho_{\infty}(K)$. Alternatively it is possible to elimitate
$\rho_\infty$ by 
using two lattice sizes. Along with  \eref{scalingfn} and
\eref{gendens} this gives a new finite-size scaling function
\begin{equation}\label{nscale}
\tilde{\varphi}_L(K)=\frac{
\log\left(\frac{\rho(K,L)-\rho(K,L-2)}{\rho(K,L-2)-\rho(K,L-4)}\right)}
{\log\left(\frac{L}{L-2}\right)},
\end{equation}
which, as before, may be used to find estimates of the critical lines
by looking for
solutions of the equation
\begin{equation}
\tilde{\varphi}_L(K_c^L)=\tilde{\varphi}_{L-2}(K_c^L).
\end{equation}
In the limit $L\to \infty$ the solutions of this equation tend to
the correct fixed points, such that $\tilde{\varphi}(K_c^L) \to 1/\nu-3$.
This again  enables the calculation of the critical exponent
$\nu$. 

\begin{figure}
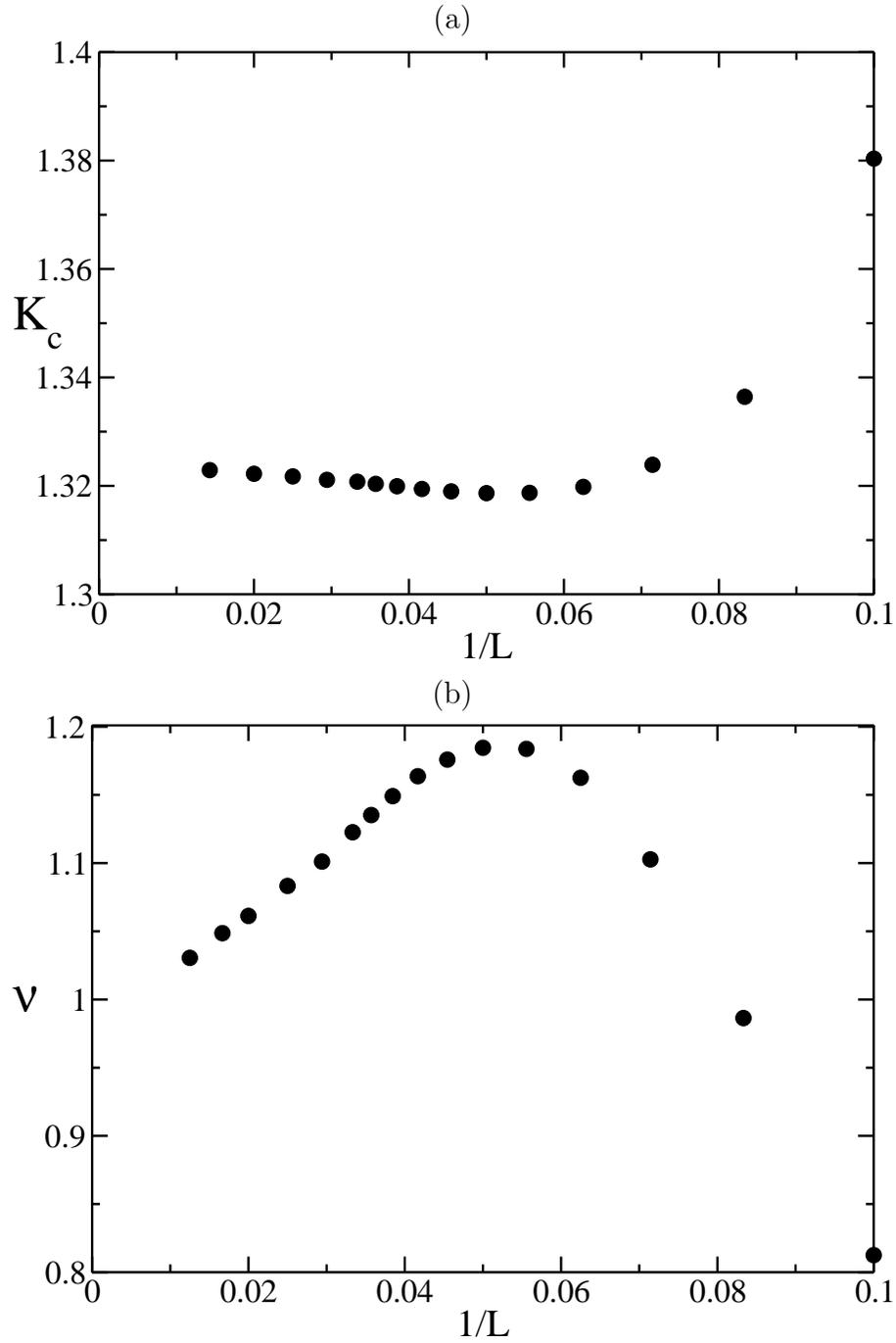

\caption{Estimates of $K_c$ and $\nu$ 
corresponding to the high-density phase transition plotted as a
function of $1/L$  
for $p=1$ and $\tau=1$. 
The estimated
value of $K_c=1.326\pm 0.002$ and convergence to the expected value of
$\nu=1$ is observed.}\label{highrho}
\begin{center}
(a)

\includegraphics[width=12cm,clip]{p1t1kc}

(b)

\includegraphics[width=12cm,clip]{p1t1nu}
\end{center}
\end{figure}

The high-density transition line is expected to be in the
Ising universality class, with an exponent $\nu=1$.  It is important
to note that the exponent is used to represent both the critical
exponent related to the correlation length   and the inverse of
the Haussdorf dimension, $d_H$\cite{Man}. The Haussdorf dimension is
the fractal 
dimension defined through the scaling of the density with the distance
from the centre of mass of the walk.
This leads to the scaling law:
\begin{equation}\label{rg}
R_G\sim N^{1/d_H},
\end{equation}
where $R_G$ is the radius of gyration, or any other characteristic
measure of the linear size of the walk.
Since in the low density regime
the gyration radius is proportional
to the correlation length, then as the critical line is approached,
and $L\to \infty$, we may identify $\nu=1/d_H$. It is common
practice to define $\nu$ from \eref{rg}, whence $\nu=1/d=1/2$ on
the first-order transition line, where the fractal and space dimension
become the same. The $\nu$ associated with the high-density transition
will necessarily not be defined in this geometrical fashion, but
corresponds to the true thermodynamic critical exponent related to the
correlation length. The estimate of $K_c$ and the associated estimates
for $\nu$ are given in \Fref{highrho} for $p=1$ and $\tau=1$. 
The evolution of $\nu$ is not
monotonic, emphasising the need to go to large enough lattice sizes
before having reliable estimates of the exponent. It is clear that
$\nu$ converges to the expected value $\nu=1$, and we estimate the
critical fugacity in this case to be
$K_c=1.326\pm0.002$.

\begin{figure}
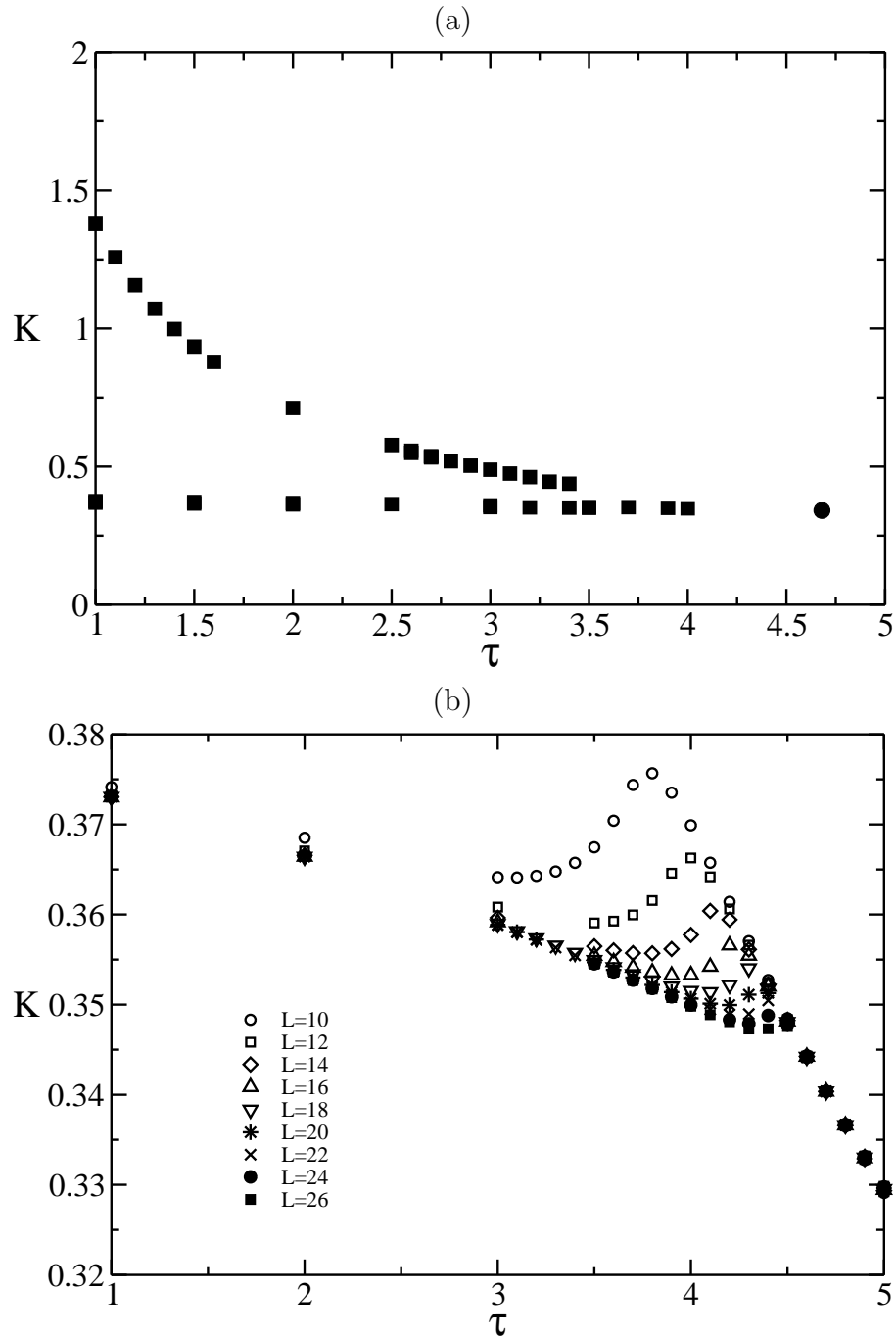

\caption{The phase diagram calculated using (a) \eref{nscale} with
$L=30$ and (b) \eref{oldsfn}, having fixed $\rho_\infty=0$, plotted
for different values of $L$ between 10 and 
26.
In (a) solutions corresponding to the low and high density phase
transitions are shown, while in (b) only the low density transition
was obtained.
 The circle in (a) shows the numerically estimated location of the
multi-critical point (see text and \Fref{p1_est}).
}\label{p=1pd} 
\begin{center}
(a)

\includegraphics[width=12cm,clip]{pdl2}

(b)

\includegraphics[width=12cm,clip]{pdl}
\end{center}
\end{figure}

\Fref{p=1pd}(a) shows the phase diagram where both transition lines
are calculated using the
scaling function given in \eref{nscale}.
In practice there are
usually two solutions to \eref{nscale} 
corresponding to the same
transition; both converge towards 
the same value of $K_c$ as the size of the lattice is
increased. However, one of these solutions evolves more slowly than
the other
as system size is increased, and this is taken as being the relevant
solution for the calculation of $\nu$ and $K_c$. As $\tau$ is
increased these two solutions come closer (for fixed lattice size),
and join at a  certain value of $\tau$. Beyond this value of $\tau$
there is nolonger a solution corresponding to the transition in
question (there may be solutions corresponding to other transitions).
This vanishing point
moves to larger $\tau$ as the lattice size is increased. The phase
diagram shown in \Fref{p=1pd}(a) is calculated for system sizes up to
$L=30$. 

\Fref{p=1pd}(b) shows
the estimates for the zero-density phase
transition lines calculated using \eref{oldsfn}. As $\tau$ approaches
its multicritical value, the finite sized system feels the presence of
the high-density phase transition, leading to a peak in the estimated
transition line. This peak reduces in size rapidly as the system size
is increased.

\begin{figure}
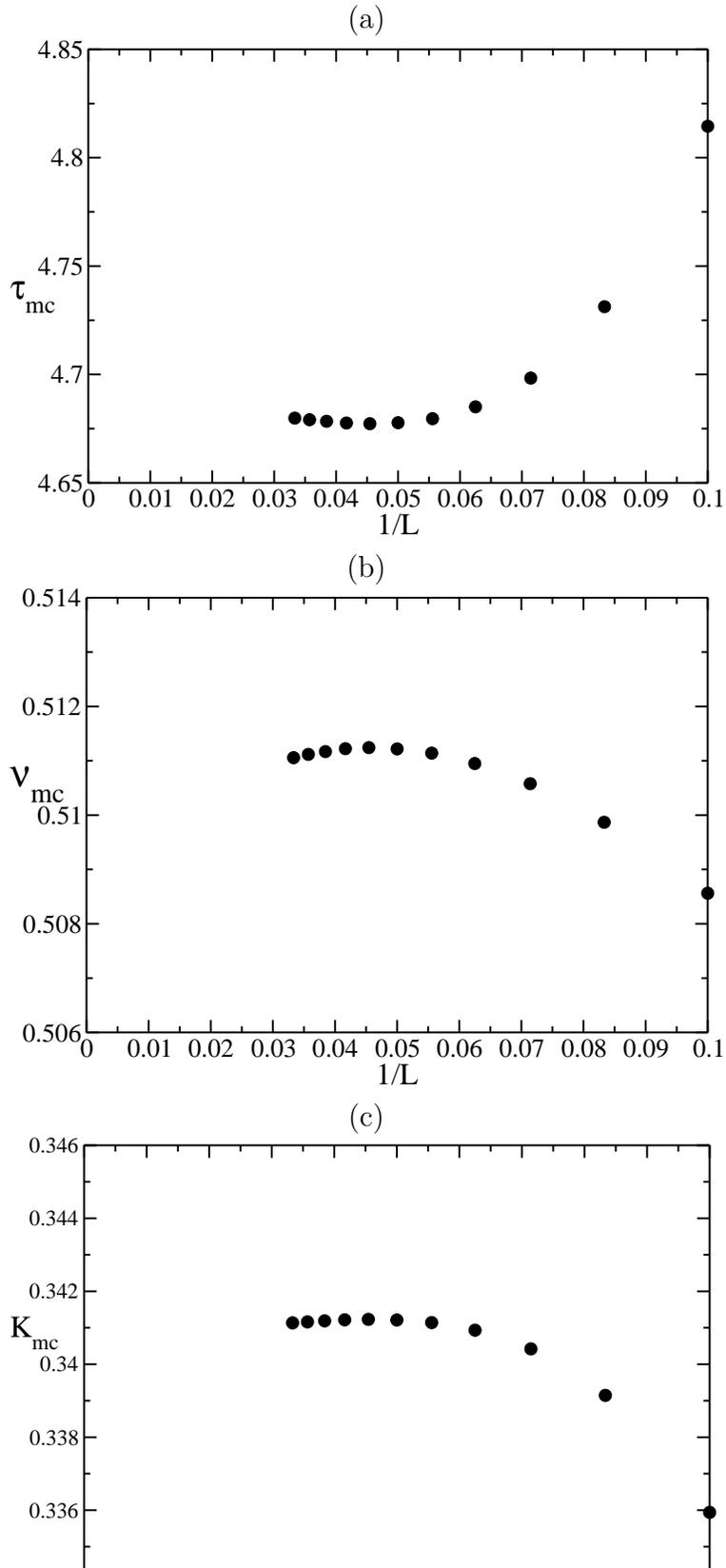

\caption{Estimates of (a) $\tau_{mc}$, (b) $\nu_{mc}$ and (c) $K_{mc}$
calculated from the intersections of $\nu_L(\tau)$ for the
vertex-interacting model with $p=1$ and fixing $\rho_\infty=0$. The
estimated position of the multicritical point is then found to be
$K_{mc}=0.3408\pm 0.0002$ and $\tau_{mc}=4.69\pm0.01$. The
estimated value of $\nu_{mc}$ is incompatible with the exact value
$\nu_{mc}=12/23\approx 0.5217$ for reasons explained in the
text.}\label{p1_est} 
\begin{center}
(a)

\includegraphics[width=10cm,clip]{p1tc}

(b)

\includegraphics[width=10cm,clip]{p1nu}

(c)

\includegraphics[width=10cm,clip]{p1kc}
\end{center}
\end{figure}

\begin{figure}
\caption{The density plotted as a function of $\tau$ calculated along
the critical self-avoiding walk line $K_c(\tau)$. The density
appears to be developing a jump at around $\tau_{mc}$ as the system
size, $L$, is increased.}\label{rhopone}
\begin{center}
\includegraphics[width=12cm,clip]{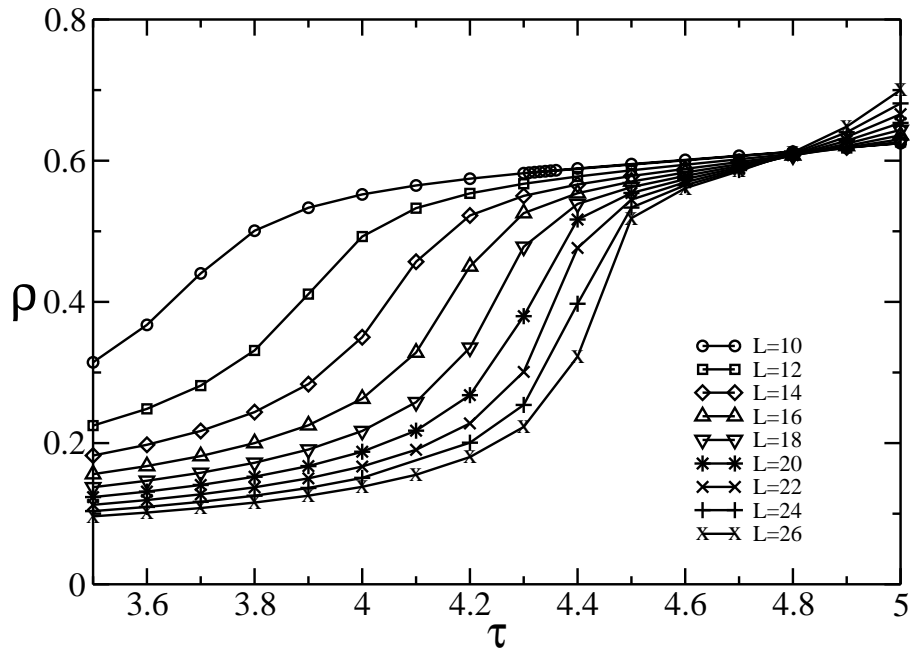}
\end{center}
\end{figure}

\begin{figure}
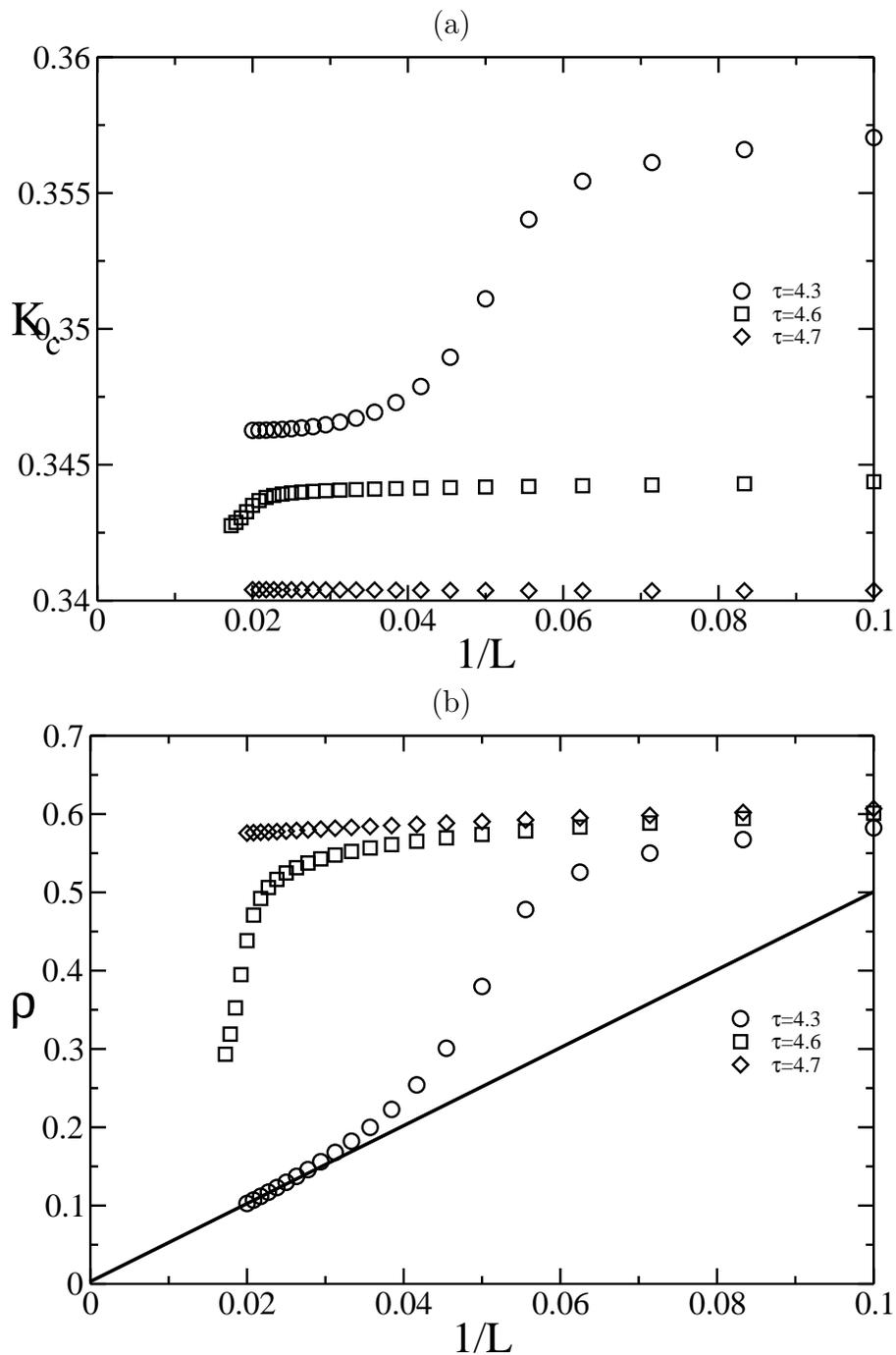

\caption{(a) Shows the estimates of $K_c$ and (b) estimates of the
density plotted as a function of 
$1/L$ for three values of $\tau$: $\tau=4.3$ and $\tau=4.6$ which are
below $\tau_{mc}$ and $\tau=4.7$ which is above $\tau_{mc}$. When
$\tau<\tau_{mc}$ the
graphs show a transit from being influenced by the high-density
transition line (small $L$) to being influenced by the zero-density
transition line (large $L$). Whilst the transit is not complete for
the sizes considered when $\tau=4.6$, it is clear that $\rho_\infty=0$
in this case whilst $\rho_\infty\neq 0$ when $\tau=4.7$. The solid
line is given as a guide to the eye.}\label{all}
\begin{center}
(a)

\includegraphics[width=12cm,clip]{allk}

(b)

\includegraphics[width=12cm,clip]{allrho}
\end{center}
\end{figure}

Estimates of the  location of the multi-critical point, ($K_{mc},\tau_{mc}$),
may again be found
by looking for crossings in the finite-size 
estimates of $\nu(\tau)$ calculated
along the self-avoiding walk transition line using \eref{oldsfn} and
\eref{ssol}. These  
estimates are given in \fref{p1_est}.
The  location of the multicritical
point is estimated as $K_{mc}=0.3408\pm 0.0002$, $\tau_{mc}=4.69\pm
0.01$. These estimates agree very well with the estimates given by
Guo, Nienhuis and Bl\"ote\cite{nien}, 
who give $z_{mc}=K_{mc}^2\tau_{mc}\approx
0.54$, however
the finite-size  estimates of $\nu_L(\tau)$ cross at a value too
low to be consistent with the exact value at the multicritical point
($\nu_{L\to\infty}\approx 0.51$ compared with $\nu_{\rm
exact}=12/23=0.5217\cdots$).  
This leads to the conclusion that either the
scaling method is flawed or $\rho_\infty\ne 0$ at the 
multicritical point. 
In  \Fref{rhopone} the density is plotted as a function of $\tau$
along the critical (self-avoiding walk) line.
At the  multicritical point the density has a clear
finite limit as the size is increased, indicating that the transition
is first order as approached in this direction. 
As further evidence of the first order character of the transition
when approached following the self-avoiding walk line we plot in
\Fref{all} the finite-size estimates of $K_c$ and $\rho$ for different
values of $\tau$ near $\tau_{mc}$. 
From \Fref{p=1pd} it is seen that for $\tau$ close
to $\tau_{mc}$ the system is influenced by the high-density
transition. As the system size is increased, the estimates transit to
the zone of influence of the zero-density transition, where they tend
to their limiting values. This transit is clearly seen in \Fref{all} for
$\tau=4.3<\tau_{mc}$. When $\tau=4.6$ it is seen that the transit begins, but
the sizes we could consider in a practical time scale were not large
enough to see the end of the transition. It is however reasonable to
suppose that for $\tau=4.6$ the limiting density will be zero and
occur as $L\to \infty$. No such transit is seen for $\tau=4.7>\tau_{mc}$, which
displays a non-zero limiting density. The precise value of this
limiting density is of course not reliable, since the location of the
transition point, found assuming $\rho_\infty=0$, will not be precise.
The
determination of the multicritical 
point is however expected to be correct since $\rho_\infty=0$ 
along the entire self-avoiding walk line, as must be the case since
the fractal dimension $d_H=1/\nu<2$.

\begin{figure}
\caption{Estimates of $K_c$ calculated using \eref{oldsfn} with 
$p=p^*_{mc}$ and $\tau=\tau_{mc}^*$ fixed at the values given 
by the exact solution~\eref{exactpt}. The estimates converge well to
the exact value $K_{mc}^*=0.446933\cdots$ (shown by the solid line) as $L\to
\infty$.}\label{ptfixed}
\begin{center}
\includegraphics[width=12cm,clip]{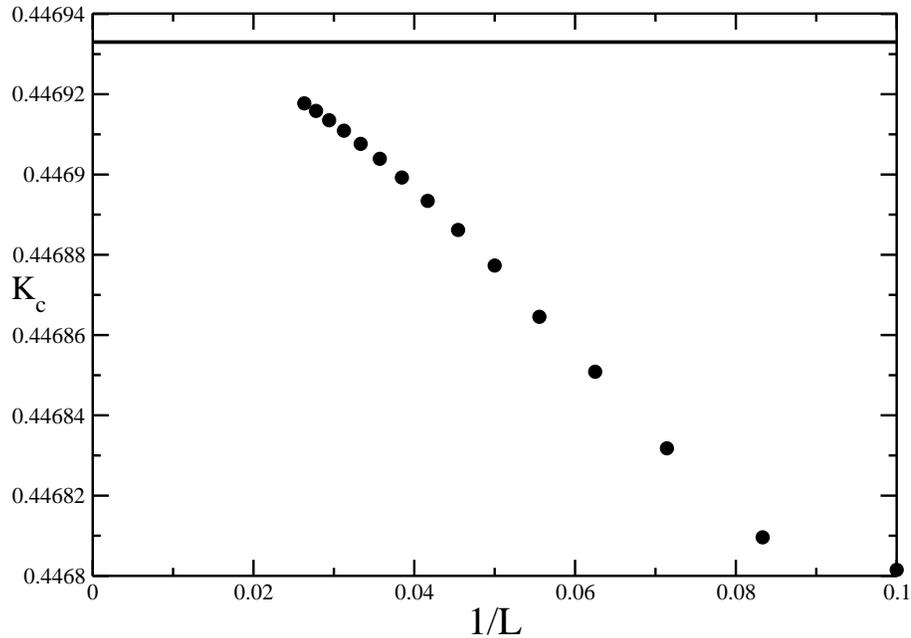}
\end{center}
\end{figure}

\begin{figure}
\caption{The density plotted for (a) $p=p^*_{mc}$ and
$\tau=\tau_{mc}^*$ given by the 
exact solution \eref{exactpt} whilst $K_{mc}$ is calculated from
\eref{oldsfn} (filled circles) and (b) $p$, $\tau$ and $K$ fixed by
\eref{exactpt}  (squares). The density does not become zero in the
limit $L\to \infty$. A conservative estimate of the limiting density
would be $0.43\pm 0.01$.}\label{rhonn}
\begin{center}
\includegraphics[width=12cm,clip]{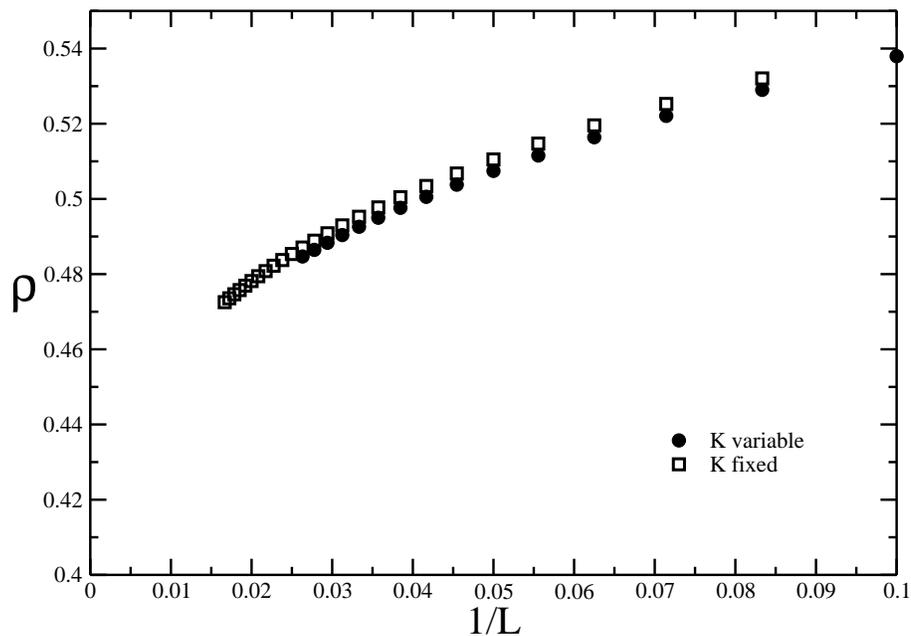}
\end{center}
\end{figure}

There exist five branches of solutions for the Bl\"ote-Nienuis $O(n)$ 
model on the square lattice\cite{nienon}, 
through its mapping onto the 19 vertex model, for which the
(multi)critical points are exactly known, as well as many of the
critical exponents. The multicritical point here is thought to
correspond to the same universality class as the $n=0$ solution
belonging to the  third branch. The location of this point is given
exactly as\cite{nienon}:
\begin{equation}\label{exactpt}
\left.\begin{array}{rcl}
z=K^2\tau&=&\left\{2-\left[1-2\sin(\theta/2)\right]
\left[1+2\sin(\theta/2)\right]^2\right\}^{-1}\\
K&=&-4z\sin(\theta/2)\cos(\pi/4-\theta/4)\\
pK&=&z\left[1+2\sin(\theta/2)\right]\\
\theta&=&-\pi/4
\end{array}\right\}.
\end{equation}
This gives the location of the exactly known multicritical point as
$K^{*}_{mc}=0.446933\cdots$, $p^{*}_{mc}=0.275899\cdots$ and
$\tau^{*}_{mc}=2.630986\cdots$. We 
checked the determination of this point using CTMRG fixing $p=p_{mc}^*$ and
$\tau=\tau_{mc}^*$ and looking for the limiting value of $K_L$. This
is shown in 
\Fref{ptfixed} and is clearly consistent with the exact known
value. The density was then calculated firstly at the finite-size
estimate of the critical point and then fixing simultaneously the
values of $K$, $p$ and $\tau$ to the
values given in \eref{exactpt} (see \Fref{rhonn}). 
In both cases a conservative estimate gives
$\rho_\infty=0.43\pm0.01$. These results indicate
that the ``multicritical'' 
point is fairly exotic in that it corresponds generically to a
first-order transition but with a critical  behaviour if approached in
a specific direction. We have tried to recover the critical exponent
$12/23$ by using the scaling behaviour of the density at $K_{mc}^*,
\tau_{mc}^*$ and $p_{mc}^*$. Whilst the results are
not inconsistent with $\nu=12/23$, they are not of very good quality
and are also not inconsistent with a value $\nu\approx 0.54$ as
originally found by Bl\"ote and Nienhuis using transfer
matrices\cite{nienon}. Whilst in principle there is no limit to either the
precision or lattice sizes
obtainable using CTMRG, there is a practical limitation in terms of
computing time and resources.

\section{Conclusion}

In this paper we have presented a version of the corner transfer matrix
renormalisation group method suitable for use with self-avoiding walk
type models and related $O(n)$ symmetric models. We have applied it to
the vertex-interacting self-avoiding walk model. The method was shown
to reproduce the known results, and has given 
new insight into the critical nature of the model when $p\neq0$.

Self-avoiding walk models are used as models for polymers in dilute
solution, and 
it is known that the
lattice $\theta$-point model provides a good description of
 what happens for real homopolymers  in solution\cite{genbks}. 
The standard textbook reason for not taking into account the rigidity
of the polymer  is that the rigidity simply changes the
persistence length, or in terms of the lattice model the number of
real monomers per step of the walk. The critical behaviour is
unaffected in the infinite walk limit.
This should be true as long as the fractal dimension $d_H$ is smaller than
the spatial dimension (i.e. $d_H<2$ here). However, when $d_H=d$ it
may be expected
that the polymer feels the presence of the lattice. This will then
give  rise to competition between the rigidity and the underlying
lattice structure, reminiscent of frustration in
spin models (indeed the critical behaviour of the
Nienhuis $O(n)$ model on the square lattice has been related to the
fully-frustrated XY model\cite{nienon}). 
If such a claim is true, then the critical behaviour observed will be
universal (in the sense of not depending on the lattice) and
independent of rigidity if $d_H<d$,
and may be non-universal (in the same restricted sense) and may depend
on the rigidity if $d_H=d$. 
One could therefore argue that the simple observation that the critical
behaviour of the collapse transition changes as $p$ is varied implies
that the density is non-zero at the collapse transition when $p\neq
0$. In interpreting this transition as due to frustration, it should  
be noted that it does not
exist for the $O(n)$ model on the hexagonal
lattice\cite{nienhuis82,nienhuis87}.  

There are other models which have 
qualitatively similar phase diagrams. Most well known is the
self-avoiding trail
model\cite{trails}. A trail is defined to be self-avoiding on the bonds, but is
allowed to visit sites more than once, and in particular is allowed to
intersect itself. The attractive interaction is again associated with
doubly visited sites (intersection or collision). 
It has been conjectured that the collapse
transition for trails occurs exactly at $\tau=3$ with an exponent 
$\nu=1/2$; it is also conjectured that limiting density is zero. 
This conjecture resulted from
a Monte-Carlo enumeration
of extremely long walks of the kinetic trail model,
believed to be equivalent to the trail model at the collapse
transition\cite{owc1995} . 
As discussed above, in this case what is being calculated is in fact
the fractal dimension $d_H$. The relation $\nu=1/d_H$ is only valid 
when the critical length may be measured using the gyration
radius. As argued above, the validity of this
identification may be questioned when $d_H=d$.
It would be of interest, then, to check by other means the possibility
that the criticality is in fact not of the same type as observed here
for $p\neq 0$. This question is currently under investigation.

Another model with a qualitatively similar  phase diagram is the
Hydrogen-bonding model\cite{fs}. 
 This model includes interactions
between nearest-neighbour sites, which may be taken into account 
in the  CTMRG method by mapping the walk model onto a 4-state vertex
model with complex weights.
There are, however, important
differences, notably 
the high-$\tau$ dense-walk phase has
$\rho\equiv 1$ throughout, whereas here $\rho\neq 1$ except in the
limit $\tau\to \infty$.
To what extent this may affect the
nature of the phase transitions is not clear; early results indicate
that perhaps the dense-walk phase transition is not in the Ising
universality class\cite{inprog}.


\pagebreak

\end{document}